\documentclass[aps,prb,preprint,showpacs,amsmath,amssymb]{revtex4}
\usepackage{subfigure}
\usepackage{graphicx}
\begin{document}

\title{Optimization-based string method for finding minimum energy path}

\author{Amit Samanta}
\email{asamanta@math.princeton.edu}
\affiliation{Program in Applied and Computational Mathematics,
  Princeton University, Princeton, New Jersey, USA.}

\author{Weinan E}
\affiliation{Department of Mathematics and
Program in Applied and Computational Mathematics, 
Princeton University, Princeton, New Jersey, USA, and\\
Beijing International Center for Mathematical Research, Peking
University, Beijing, China.}
\date{\today}

\begin{abstract}
We present an efficient algorithm for calculating the minimum energy
path (MEP) and energy barriers between local minima on a
multidimensional potential energy surface (PES). Such paths play a
central role in the understanding of transition pathways between
metastable states. Our method relies on the original formulation of
the string method [Phys. Rev. B ${\bf 66}$, 052301 (2002)], i.e. to evolve a
smooth curve along a direction normal to the curve. The algorithm
works by performing minimization steps on hyperplanes normal to the
curve. Therefore the problem of finding MEP on the PES is remodeled as
a set of constrained minimization problems. This provides the
flexibility of using minimization algorithms faster than the steepest
descent method used in the simplified string method [J. Chem. Phys.,
${\bf 126}$(16),164103 (2007)]. At the same time, it provides a
more direct analog of the finite temperature string method.  The
applicability of the algorithm is demonstrated using various
examples. 
\end{abstract}

\maketitle
\section{Introduction}
The dynamics of complex systems often involve thermally activated
barrier-crossing events that allow the system to move from one local
minimum of the energy surface to another. At finite temperatures, the
total kinetic energy accessible to the system is on the order of
$Nk_{\rm B}T$, where $N$ is the number of degrees of freedom, $k_{\rm
  B}$ is the Boltzmann constant and $T$ is the temperature of the
system. However, this huge amount of energy is distributed over the
whole system. Consequently, it fails to cross over the free energy
barrier (generally an index-1 saddle point) and move to a different
basin of attraction. A system can overcome a free energy barrier only
when sufficient energy is localized on an activated region of the
system. The activated region is the volume of the sample where bond
breaking/formation, atomic re-arrangements etc. take
place.$\cite{DaoLADM07, AsaroS05}$ It is of great theoretical and
practical interest to develop algorithms that can enable us to
efficiently compute the most probable pathways for 
such transition events. For systems with relatively smooth energy 
landscapes, it can be shown that the most probable pathways are the
minimum energy paths (MEP). Minimum energy paths are physically
relevant in the low temperature dynamics of a system and provides
information only about the energy barrier involved in a thermally
activated event without any consideration of the width of the channel
near the saddle point or other entropic effects. Further, at high temperature,
the energy surface becomes rugged due to thermal fluctuations and the
presence of multiple peaks of $\mathcal{O} \left(k_{B}T \right)$ makes
the concept of MEP irrelevant. However in such a scenario the MEP can
still correspond to the path with the maximum likelihood. 

The problem of finding MEP and the bottlenecks for transition events
can be broadly categorized into two classes depending on the initial
conditions: (a) when only the initial point is known, and (b) when
both the initial and final points on the energy surface are available. 
In the former case, one can resort to methods like gentlest ascent
dynamics$\cite{EX11}$, dimer method$\cite{Henkelman99}$, etc. to
explore the energy surface. For the second category, the most notable
examples include the string method$\cite{ERV02, ERV07, MaraglianoV07,
  Vanden-Eijnden09}$ and the nudged elastic band
method$\cite{HenkelmanJ00, HenkelmanUJ00, SheppardTH08}$. In this
case, we are given the initial and final states of the system, and our
aim is to find the MEP connecting these states. Since
there can be multiple paths joining the end points, the converged MEP
is dependent on the choice of the initial path. In the original
string method, a path $\gamma$ evolves as:
\begin{equation}
  \dot{\gamma} = -\nabla V\left(\gamma\right)^{\perp} + r\hat{t}
\end{equation}
where, $\dot{\gamma}$ is the time derivative of $\gamma$, $\nabla
V^{\perp} = \nabla V - \left(\nabla   V,\hat{\tau}\right)\hat{\tau}$
is the gradient of the potential perpendicular to $\gamma$,
$\hat{\tau}$ is the unit vector parallel to the tangent to $\gamma$
and $r$ is a Lagrange multiplier used to enforce a particular
parametrization of the string. 

The string method is easy to implement and works well even if the
initial and the final states are not local minimum.$\cite{ERV02}$ This
advantage has led to a class of algorithms called growing string
method that have been tested for real systems
using quantum mechanical tools.$\cite{PetersHBC04, Quapp05,
  QuappKC07}$ Further, if one is solely interested in knowing the free
energy barrier and the configuration at the saddle
point, the end of the string need not be a local minimum
but some intermediate configuration lying in a basin other than that
of the initial point.

The string method is based on the idea of
moving curves by using a steepest descent-type of dynamics. It should
be emphasized that even though the dynamics used in the string method
has very strong steepest decent flavor, the method does not amount
to minimizing any energy functionals. In order to apply quasi-Newton
type of ideas to accelerate the string method, one has to resort to
the Broyden formulation for solving coupled system of equations rather
than the BFGS type of formulation for
optimization.$\cite{WeiQingRenThesis}$ 

We propose a method that reduces the problem of finding MEP to an
optimization problem, so that techniques from optimization theory can
now be used directly. This prescription is in some ways an improved
version of the locally updated planes (LUP) method proposed earlier
by Elber and co-workers.$\cite{CzerminskiE89, ChoiE91}$ In the LUP
method, each intermediate configuration along the approximate path is relaxed
by confining its motion to a hyperplane. The hyperplanes are selected
such that they are perpendicular to the straight line joining the two
end points. The LUP method is unstable because - (i) each intermediate
configuration is
relaxed independently and the algorithm does not prescribe any scheme
to make sure that the path is smooth, (ii) the hyperplane selection
scheme can result in the formation of kinks if the path has
multiple local energy wells, and (iii) since there is no prescribed
way to control the separation between the intermediate configurations,
the relaxed path can have images clustered around the local
minima. These problems severely limit the convergence to the correct
MEP in the LUP method.

The finite temperature string (FTS) method is like
an expectation-maximization (EM) algorithm for curves.$\cite{WeinanRV05,
  Vanden-EijndenV09}$ The current version of the string method is more
in line with the spirit of FTS in which the sampling procedure is
replaced by the minimization step. 

\section{The method}

Given a curve $\gamma$ that joins two points ${\bf{A}}$ and ${\bf{B}}$
on the energy surface, let us parametrize $\gamma$ as $\gamma =
\left\{{\bf X}\left(\alpha\right):\;\alpha\in\left[0,1\right]\right\}$
  where $\alpha$ is a continuous parameter. When the end points ${\bf{A}}$ and
  ${\bf{B}}$ lie in different basins of attraction, the curve $\gamma$
  will pass through multiple saddle points that are generally the
  bottlenecks of transition of a system from ${\bf{A}}$ to
  ${\bf{B}}$. The path $\gamma$ is the MEP if
\begin{equation}
  \left({\bf{\nabla}}{V}\right)^{\bot}
  \left({\bf{X}}\left(\alpha\right)\right) = 0
  \label{MEP1}
\end{equation}
where $(\nabla V)^{\bot} = \nabla V - (\tau, \nabla V) \tau$ is the
component of $\nabla V$ normal to the path $\gamma$ and $\tau$ is the
unit tangent vector of $\gamma$. If ${\bf{P}}\left(\alpha\right)$ is
the hyperplane perpendicular to the path $\gamma$ at
${\bf{X}}\left(\alpha\right)$, then ($\ref{MEP1}$) can be restated as
\begin{equation}
  {\bf{X}}\left(\alpha\right) = {\rm argmin}\left.{V}\right|_{{\bf{P}}\left(\alpha\right)}
\label{MEP2}
\end{equation}
This is a variational characterization of the MEP. The new version of
the string method is based on the variational formulation
in $\left(\ref{MEP2}\right)$. The initial guess path is first discretized
to give $\left\{{\bf{X}}_{j}^{0}\right\}_{j=1, \cdots, N}$. At step
$m$, the path is updated by the following procedure: 
\begin{enumerate}
\item {\it Minimization}: a number of minimization steps are performed on
  the hyperplanes normal to the curve at the discretization
  points. This gives $\left\{{\bf{X}}_{j}^{*}\right\}_{j=1, \cdots, N}$.
\item {\it Mixing}: a mixing scheme is applied to get 
  \begin{equation}
    \tilde{\bf{X}}_{\rm j} = {\bf{X}}_{\rm j}^{\rm m}\left(1-\lambda\right) + 
    \lambda{\bf{X}}_{\rm j}^{\ast}
  \end{equation}
  where, $\lambda\in\left(0,1\right)$ is the mixing co-efficient. 
  This step helps in stabilizing the scheme.
\item {\it Reparametrization}: redistribute the intermediate configurations
  along the string according to some metric, such as, equal spacing in
  configuration space, or equal spacing in energy space, etc.
  \begin{equation}
    \left\{\tilde{\bf{X}}_{\rm j}\right\} \overset{\rm reparametrize}
    {-\hspace{-0.5em} -\hspace{-0.5em}-\hspace{-0.5em}\longrightarrow} 
    \left\{{\bf{X}}_{\rm j}^{\rm m+1}\right\}
  \end{equation}
\end{enumerate}
These steps are performed until convergence. Step 3
is a standard procedure in the string
method and accounts for the displacements parallel to the
curve. Steps 1 and 2, are different from the original and simplified
string methods.

\section{Algorithmic details}
{\bf Step 1: Minimization}\\
The minimization procedure can be implemented in different ways. The
energy of each intermediate configuration can be minimized on their
respective hyperplanes till convergence or the minimization can be
performed only for few steps. Minimization algorithms with better
convergence like, FIRE (fast inertial relaxation
engine)$\cite{BitzekKGMG06}$, conjugated
gradient$\cite{NocedalNOpt00}$, limited memory
BFGS$\cite{NocedalNOpt00}$, etc. can be used for this purpose. For the
sake of completeness below we present the modified versions of BFGS
and FIRE algorithms:\\  
{\it BFGS:} Starting with an intermediate configuration at ${\bf x}_{0} = {\bf
  X}_{j}^{0}$ on the path and an approximate Hessian matrix ${\bf
H}_{0} = \nabla\nabla V\left({\bf x}_{0}\right)$ at ${\bf x}_{0}$, the
BFGS scheme involves the following steps until convergence is achieved
:$\cite{NocedalNOpt00}$\\  
({\it i}) obtain a direction ${\bf p}_{k}$: ${\bf
  H}_{k}{\bf p}_{k} = \nabla{\bf F}\left({\bf
    x}_{k}\right)$, where ${\bf F}\left({\bf x}_{k}\right) =
-\nabla V\left({\bf x}_{k}\right)$. This is performed by obtaining
the inverse of ${\bf H}_{k}$ by applying the Sherman-Morrison scheme.\\
({\it ii}) obtain step size $\alpha_{k}$ along ${\bf p}_{k}$ (line
search)\\
({\it iii}) update image:
${\bf x}_{k+1} = {\bf x}_{k} + \alpha_{k}{\tilde{\bf p}}_{k}$,
where ${\tilde{\bf p}}_{k} = {\bf p}_{k} - \left({\bf p}_{k}, {\bf
    \tau}_{j}\right){\bf \tau}_{j}$ is the projection of the search
direction perpendicular to the tangent (${\bf \tau}_{j}$) to the
path at ${\bf X}_{j}^{0}$.\\
({\it iv}) set ${\bf s}_{k}=\alpha_{k} {\bf p}_{k}$ and ${\bf
  y}_{k} = -\left[\nabla {\bf F}\left({\bf x}_{k+1}\right) -
\nabla{\bf F}\left({\bf x}_{k}\right)\right]$.\\  
({\it v}) update Hessian:
\begin{equation}
  H_{k+1} = H_{k} +
  \frac{\mathbf{y}_{k}\mathbf{y}_k^{\mathrm{T}}}{\mathbf{y}_{k}^{\mathrm{T}}
    \mathbf{s}_k} - \frac{H_k \mathbf{s}_{k} \mathbf{s}_{k}^{\mathrm{T}}
    H_{k}}{\mathbf{s}_{k}^{\mathrm{T}} H_{k}\mathbf{s}_{k}}.\\
\end{equation}
{\it FIRE:} Starting with an initial intermediate configuration at ${\bf x} = {\bf X}_{j}^{0}$
on the path, the following dynamical system can be used for the
constrained minimization:$\cite{BitzekKGMG06}$
\begin{equation}
  \begin{split}
    {\bf \dot{x}} &= {\bf v} - \left({\bf \tau}_{j}, {\bf
        v}\right){\bf \tau}_{j},\qquad \\ 
    {\bf \dot{v}} &= \frac{1}{m}{\bf F} - \beta{\bf v} + 
    \beta\left|\bf v\right|\hat{\bf F}\\
  \end{split}
  \label{FIRE1}
\end{equation}
where, $\tau_{j}$ is the tangent to the path at ${\bf X}_{j}^{0}$, $m$
is the mass and $\beta$ is a parameter. For the intermediate images,
the tangent $\tau_{j}$ at ${\bf X}_{j}^{0}$ can be approximated as 
\begin{equation}
  \tau_{j} = \frac{{\bf X}_{j+1}^{0} - {\bf X}_{j-1}^{0}}{\left|{\bf
        X}_{j+1}^{0} - {\bf X}_{j-1}^{0}\right|},\qquad j = 2, 3, ... \left(N-1\right)
\end{equation}

{\bf Step 2: Mixing}\\
During minimization, since each intermediate image is relaxed
independently, their relaxation step lengths can vary due to which the path
develops kinks and is no longer smooth. This makes the tangent vector
inaccurate leading to numerical instability. To overcome this
difficulty, in the second step, we use a mixing scheme to have better
control over step lengths. 

{\bf Step 3: Reparametrization}\\
Next, reparametrization is performed in two steps: computing
the values of the parameter $\tilde{\alpha}$ and performing
interpolation to find the reparametrized intermediate configurations. For parametrization
by equal arc length we first obtain the length of the string:
\begin{equation}
  L_{j} = L_{j-1} + \left|\tilde{\bf X}_{j} - \tilde{\bf X}_{j-1}\right|
\end{equation}
where, $L_{0} = 0$ and $j = 1, 2,\;..., N$. The corresponding
normalized parameter is then given by $\tilde{\alpha}_{j} =
L_{j}/L_{N}$. Next, using cubic spline interpolation scheme,
we obtain the intermediate configuration positions $\left\{{\bf{X}}_{\rm j}^{\rm
    m+1}\right\}$ that are evenly distributed along the string
according to the updated parameter $\alpha_{j} = j/N$. For computational
efficiency, reparametrization may not be enforced at each
step. In fact, we have found that reparametrization is much less
important for this version of the string method than the original
version.$\cite{ERV02, ERV07}$ Thus, by
constraining the minimization on the normal hyperplanes, we have
already removed the strong tendency for the intermediate positions to
move towards the local minima.  So reparametrization only plays a minor role here for
improving the accuracy by optimizing the distribution of the points
along the curve.

Naturally one is interested in a comparison of the performance of this
version of the string method with the older version.  The answer is
that it depends on the problems one needs to deal with. Suffice to say
that the current version provides an alternative. In addition, this
current version  of the string method is much closer in spirit to the
finite temperature string method in which the minimization step is
replaced by a sampling step.$\cite{WeinanRV05}$ 

\section{Illustrative Examples}
\subsection{Two-dimensional potential}

\begin{figure}[thbp]
  \centering
  \includegraphics[width=0.5\textheight]{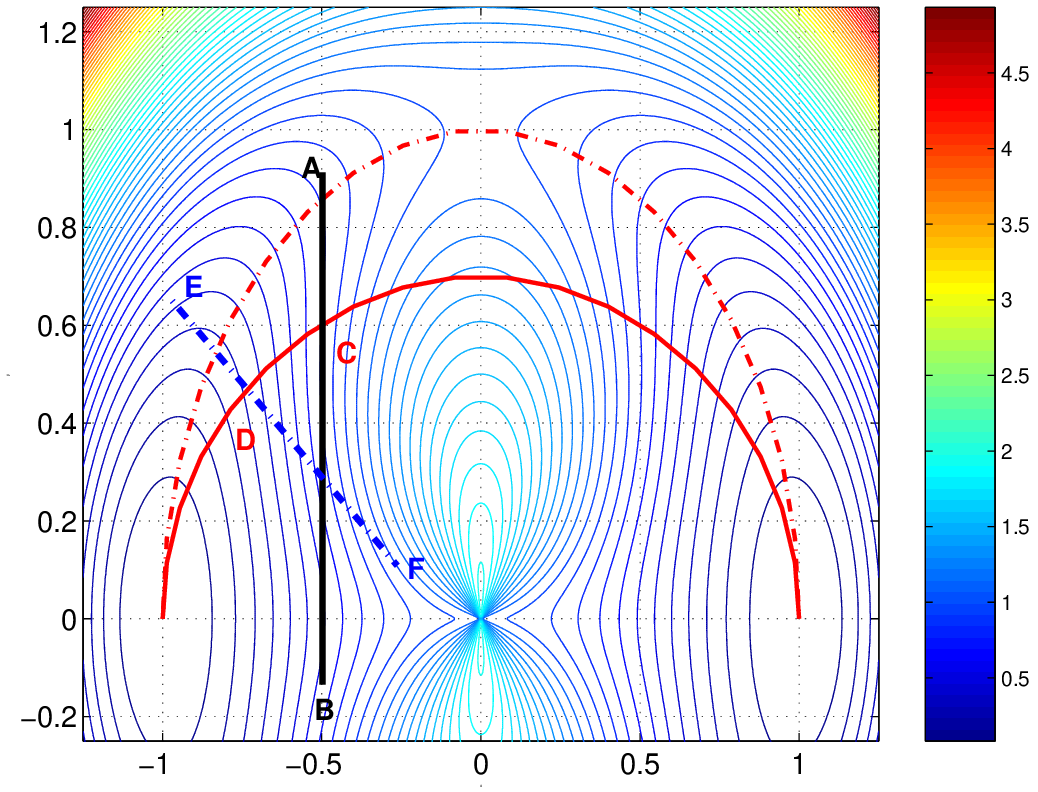}
  \caption{Contour plot of the two-dimensional potential energy
    landscape showing the MEP (red-dashed curve) and an unrelaxed path
    (red curve) joining the two local minimum points (-1,0) and
    (1,0). Lines AB and EF show the different possibilities of
    selecting hyperplanes at points C and D, respectively, on the
    string.} 
  \label{2Dmep1a}
\end{figure}

To get a better understanding of the effectiveness of the selection of
hyperplanes, let us look at the potential energy surface of a simple
two-dimensional toy problem shown in Fig. $\ref{2Dmep1a}$. The
potential energy is given by$\cite{ERV07}$
\begin{equation}
  V\left(x,y\right) = \left(1-x^{2}-y^{2}\right)^{2} +
  \frac{y^{2}}{x^{2}+y^{2}} 
  \label{Algo4}
\end{equation}
The line AB is perpendicular to the line joining the two local minimum
points (-1,0) and (1,0). C is at the intersection of the initial guess
path and AB. If we perform a constrained relaxation of C along AB,
there are more than one local minimum points in the
vicinity. Consequently, during subsequent iterations the intermediate
configuration can flip between these locally stable structures. This will make
the string uneven leading to the development of kinks due to which it can not
converge to the MEP. 

However, consider now the line EF which is normal to the path at
D. Now the point D has only one choice to lower the energy - go from D
towards E, because going towards F means going uphill which increases
the energy. Clearly, constrained minimization along hyperplanes normal
to the string is a better choice. After relaxation we obtain a smooth
string joining the two local minimum points which is shown in
Fig. $\ref{2Dmep1a}$ (red dashed curve). 

\subsection{Ad-atom diffusion on (111) surface of Cu} 
Next, we use the optimization-based string method to find the MEP
corresponding to rare events taking place in nature. As an example, we study the
diffusion of an ad-atom on the $\{$111$\}$ surface of copper. The
$\{$111$\}$ surface is used as it has lowest surface energy in Cu. The
inter-atomic interactions are modeled using embedded atom method (EAM)
potential developed by Mishin et al.$\cite{MishinMPVK01}$ 

On a pure $\{$111$\}$ surface there are multiple sites at which the potential
energy is locally minimum. One of them is the face center cubic (fcc)
hollow site and the other being the hexagonal close-packed (hcp)
hollow site. In a fcc crystal,
a set of three $\{$111$\}$ planes forms the stacking sequence while in
hcp a set of two $\{$111$\}$ planes forms the stacking
sequence. Hence, an ad-atom occupying a surface site which corresponds
to fcc (hcp) stacking is called fcc (hcp) hollow site. 

The simulation cell contains 512 atoms with cell dimensions of
37.57$\times$20.45$\times$17.71 ${\rm\AA^{3}}$ and cell axes parallel
to [111], [1$\bar{1}$0] and [11$\bar{2}$] directions. Periodic
boundary conditions are imposed along the [1$\bar{1}$0] and
[11$\bar{2}$] directions and free surface conditions are imposed along
the [111] direction. The end points of the string are configurations
with ad-atoms in hcp hollow sites separated by about 10 $\rm\AA$ on
the (111) surface. 

\begin{figure}[thbp]
  \centering
  \includegraphics[width=0.5\textheight]{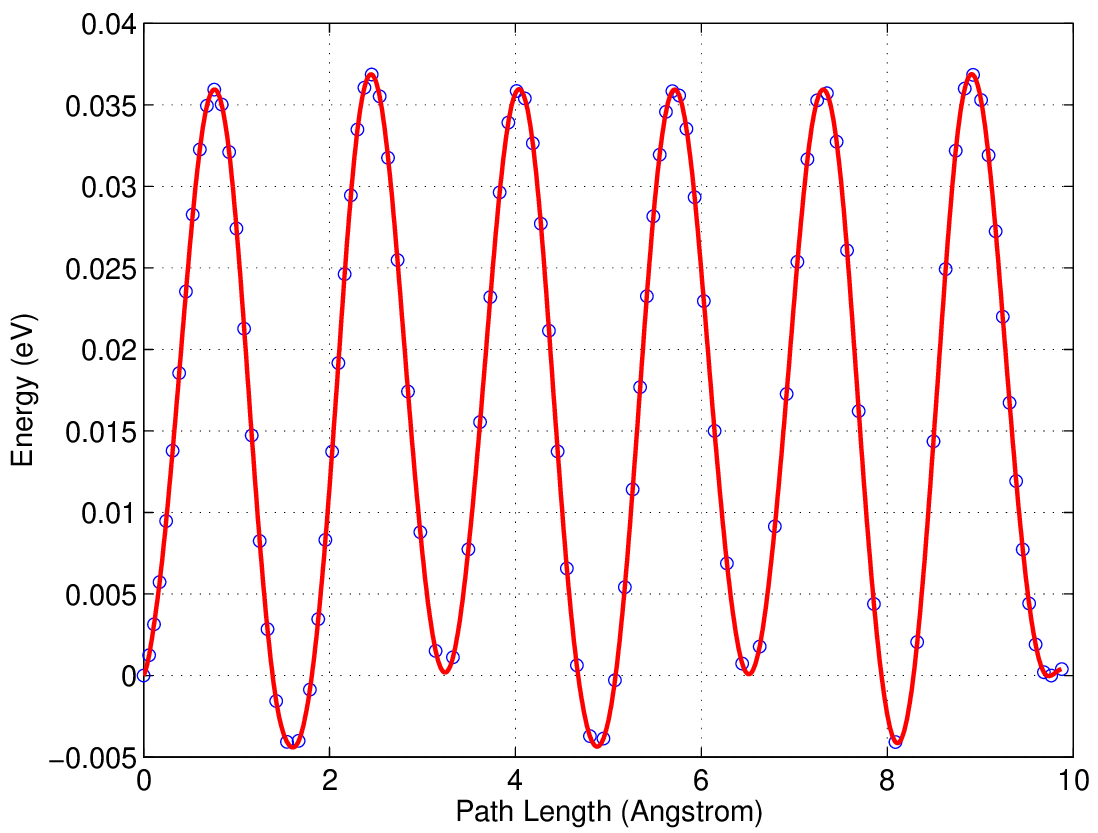}
  \caption{Converged MEP of the ad-atom diffusion on a (111) free
    surface in Cu. At the end point the ad-atom occupies hcp hollow
    sites that are about 10 $\rm\AA$ apart. The energy barrier for
    diffusion from hcp to fcc hollow site is 0.036 eV and from fcc to
    hcp hollow site is 0.04 eV. The converged MEP is shown in the
    red-curve while the blue circles show an intermediate path without
    parametrization.} 
  \label{AdatomDiffusion512}
\end{figure}

Fig. $\ref{AdatomDiffusion512}$ shows the converged MEP (force norm
less than 10$^{-3}$ eV/$\rm\AA$) as a function of the length of the
path in the multidimensional configuration space. The MEP is obtained
by performing constrained conjugated gradient minimization on each
intermediate configuration. The red curve corresponds to the
converged MEP with intermediate configurations evenly distributed (now shown
in the figure) along the path. The blue circles denote an intermediate
path without reparametrization. The intermediate structures in this case
cluster around each other. As shown in the figure, the ad-atom in hcp
site is not the lowest energy, the structure with ad-atom occupying
the fcc hollow site has lower energy ($\sim$0.004 eV). The diffusion
barrier from hcp hollow site to fcc hollow site is 0.036
eV. Similarly, the diffusion barrier from fcc to hcp hollow site is
about 0.04 eV. The results shown are without any mixing scheme to
update the image positions, i.e. $\lambda$ = 1.  

\subsection{Dislocation nucleation in a nano-wire}

Understanding the process of dislocation nucleation and multiplication
is central to our understanding of structural stability of
materials. A dislocation is a line defect which corresponds to the
interface between the sheared and unsheared regions of a sample. A
bulk sample in nature generally has an astronomical number of defects,
of varied dimensionality, present in them. In contrast, in
materials of small volume the density of defects can be very
small. Small volume materials, such as, quantum dots, nano-wires, thin
films, nanostructured materials, etc. have much higher surface area
(or grain boundary area) to volume ratio than their bulk 
counterparts$\cite{HemkerN08, Yip98}$ and defect nucleation from the
surface becomes an important driving force for plastic
deformation. Given the contrasting length-scales involved, the
deformation mechanisms changes from bulk-dominated plasticity to
surface-dominated plasticity with decrease in characteristic length
scales. One such potentially important form of deformation in
nano-wires, nano-pillars, etc. that can play a critical role in
controlling plastic deformation is dislocation
nucleation.$\cite{RabkinS07, HydeEF05, ShanMAWM08}$ Here, we perform
atomistic simulations of heterogeneous dislocation nucleation in a Cu
nano-wire. We use a nano-wire of square cross-section to include the
corner as a preferential nucleation site.

\begin{figure}[thbp]
 \centering
 \subfigure[] { \includegraphics[width=0.3\textheight]{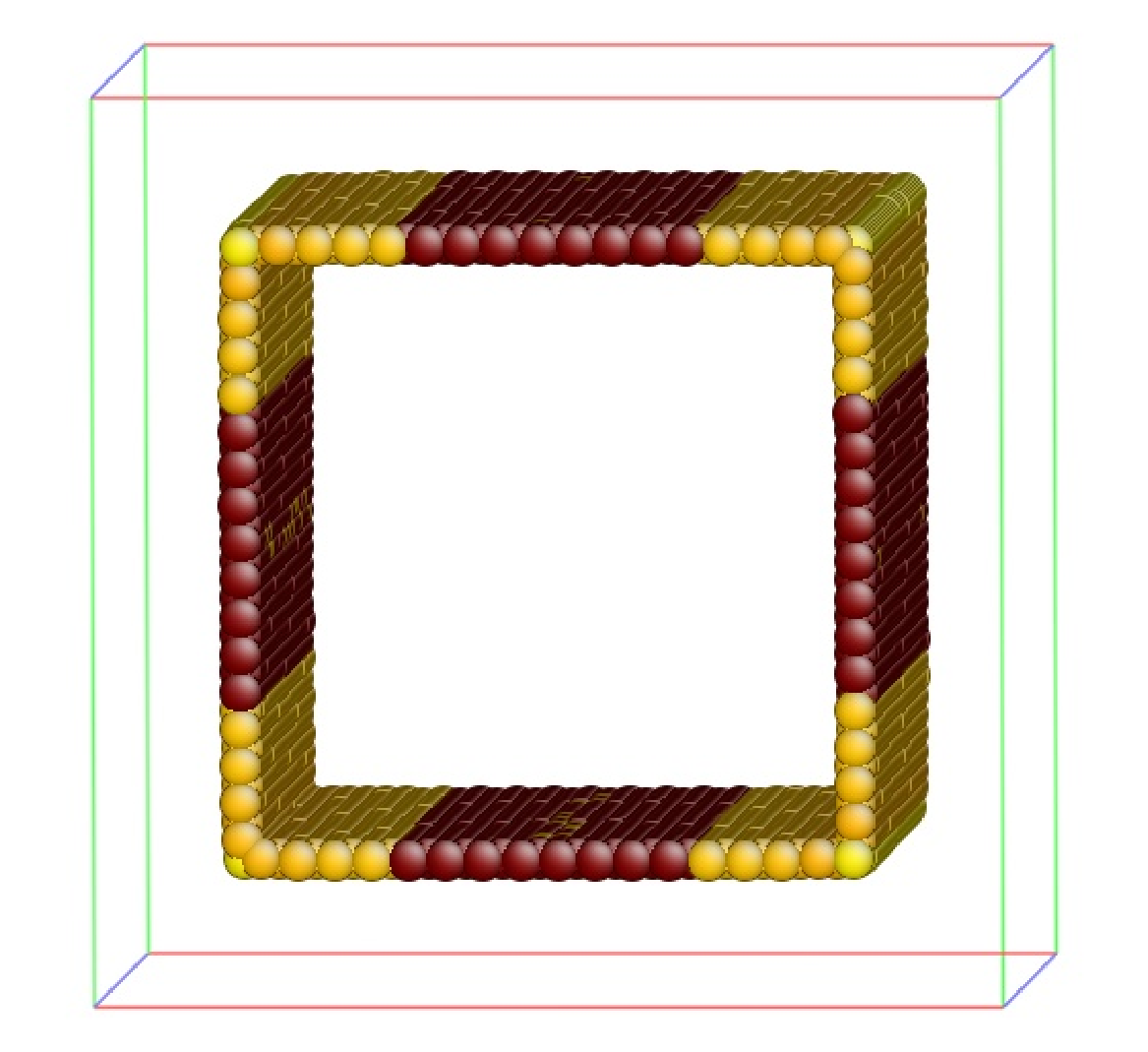} 
   \label{Dislocation1}}
 \subfigure[] { \includegraphics[width=0.3\textheight]{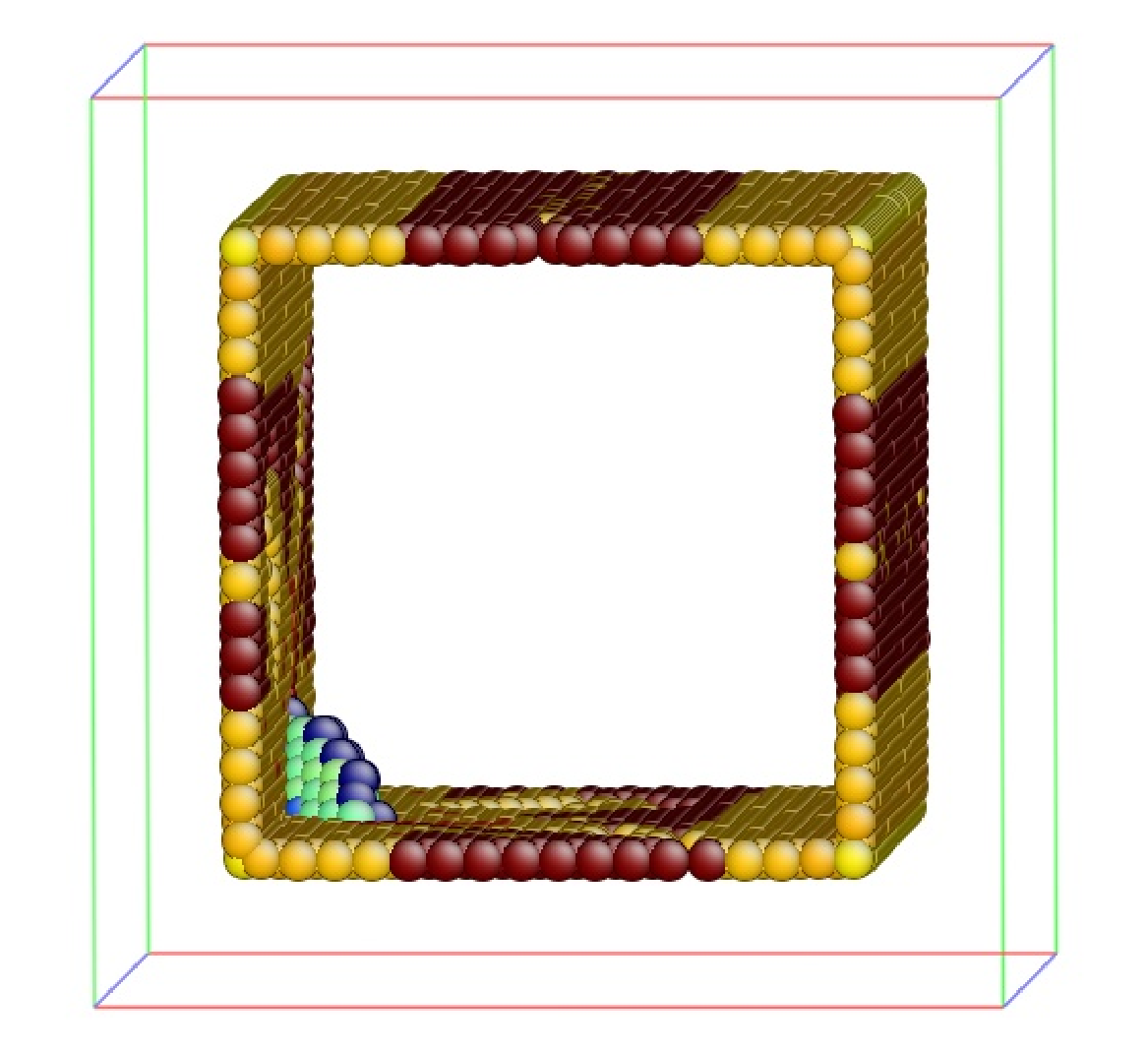} 
   \label{Dislocation2} }
 \caption{ Snapshots of the initial and the saddle configuration for
   the process of heterogeneous dislocation nucleation from the corner
   of a nano-wire. The nano-wire is under uniaxial compression of 6
   $\%$, the strain axis being parallel to the wire axis. The central
   symmetry scheme of coloring$\cite{KelchnerPH98}$ is used:
   $\ref{Dislocation1}$ only the atoms on the surface are visible,
   $\ref{Dislocation2}$ the surface and the atoms in the sheared
   region, in the saddle point configuration, are visible. The viewing
   direction is parallel to the wire axis.}  
 \label{DislocationNucleationCfg}
 \end{figure}

The simulation cell has 28,900 atoms and the cell dimensions are
90.375$\times$90.375$\times$90.375 $\rm\AA^{3}$. The cell axes are
parallel to [100], [010] and [001] directions. Periodic boundary
conditions are maintained along [001] direction (i.e. wire axis) and
free surface conditions are imposed along the remaining two
directions. The interatomic interactions are modeled using embedded
atom method (EAM) potential developed by Mishin et
al.$\cite{MishinMPVK01}$ Fig. $\ref{Dislocation1}$ shows the initial
structure of the nano-wire. Central symmetry scheme of coloring is
used.$\cite{KelchnerPH98}$ The converged MEP (force norm smaller than 
10$^{-3}$ eV/$\rm\AA$) is obtained by performing constrained
conjugated gradient minimization on each intermediate configuration
along the path. 

The initial structure is a defect-free pure Cu nano-wire and the final
structure is one with a dislocation loop (not a local minimum). The
initial structure 
is generated by applying a compressive strain of 6 $\%$ along the
nano-wire axis and then performing conjugated gradient relaxation. The
final structure with a dislocation loop is generated by applying a
relative displacement equal to a partial Burgers vector between two
$\{$111$\}$ planes. The initial guess path is generated using linear
interpolation between the end point configurations. 

\begin{figure}[thbp]
  \centering
  \includegraphics[width=0.5\textheight]{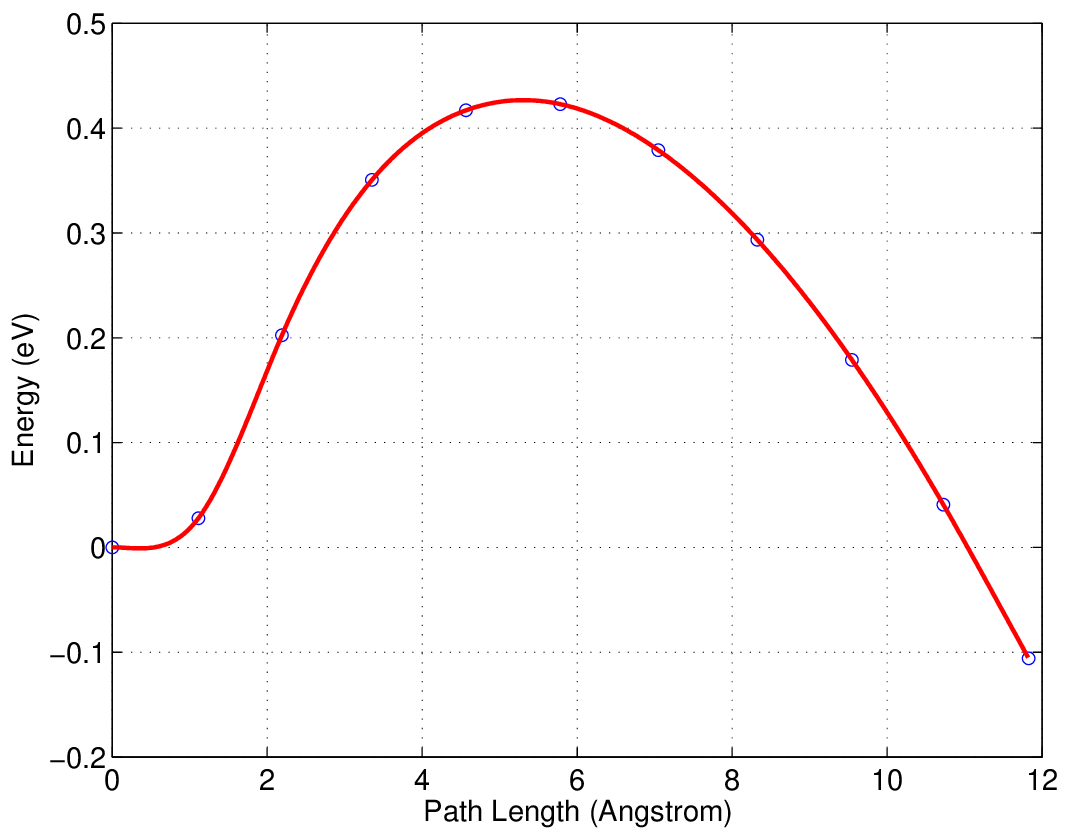}
  \caption{The converged MEP of the heterogeneous dislocation
    nucleation from the corner of a Cu nano-wire. The zero temperature
    energy barrier is 0.43 eV for a 6 $\%$ uniaxial compressive strain
    applied along a direction parallel to the wire axis. The initial
    state is a defect free Cu nano-wire with (100) surfaces.}  
  \label{DislocationNucleationCu}
\end{figure}

Fig. $\ref{DislocationNucleationCu}$ shows a section of the multi-
dimensional PES as a function of the distance between the initial
point and intermediate configurations. Fig. $\ref{Dislocation2}$ 
shows an intermediate configuration close to the saddle point. 
The activation energy barrier is 0.43 eV and the sheared region consists 
of around 30 atoms. Central symmetry scheme of coloring is used so only 
the atoms in the sheared region are visible.$\cite{KelchnerPH98}$ More 
details about the thermally activated nature of dislocation nucleation
can be found elsewhere.$\cite{ZhuLSLG08}$ 

\section{Conclusion}
In this paper a modified version of the string method is proposed. The
proposed method allows determining the MEP by finding the minimum
energy configuration along the hyperplanes normal to the path. The
algorithm is simple, easy to implement and provides the flexibility
of using faster minimization algorithms.

The applicability of the algorithm is demonstrated using a simple
two-dimensional potential energy landscape. The algorithm can be
easily extended to study systems of higher dimensions. As examples
taken from nature, diffusion barriers of an ad-atom on the surface of
Cu and the process of heterogeneous dislocation nucleation from the
corner of a nano-wire are evaluated. For the ease of reference, codes
for some sample problems are placed in a publicly accessed web
link.$\cite{StringWebLink}$ 

\begin{acknowledgments}
We acknowledge support by the Department of Energy under Grant
No. DE-SC0002623. We thank Dr. Xiang Zhou for useful discussions and
comments.
\end{acknowledgments}

\bibliography{MyBibliography}

\clearpage
\begin{figure}[thbp]
  \centering
  \includegraphics[width=0.55\textheight, angle = 270]{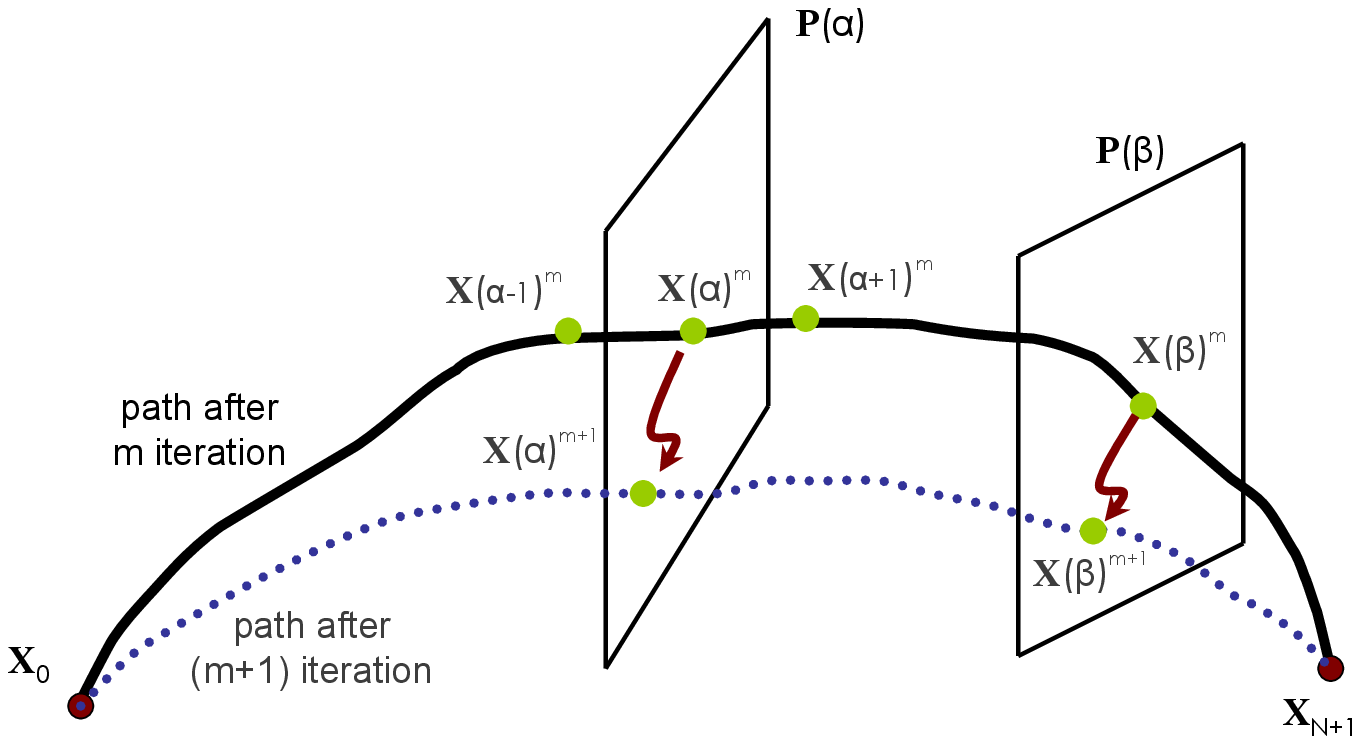}
  \caption{Schematic representation of the optimization-based string
    method to find MEP. The intermediate configuration at
    ${\bf{X}}^{m}\left(\alpha\right)$ is relaxed to ${\bf{X}}^{m+1}$
    on a plane ${\bf{P}}\left(\alpha\right)$ which is perpendicular to
    the tangent at ${\bf{X}}^{m}\left(\alpha\right)$.} 
  \label{StringFig}
\end{figure}

\begin{figure}[thbp]
  \centering
  \includegraphics[width=0.5\textheight]{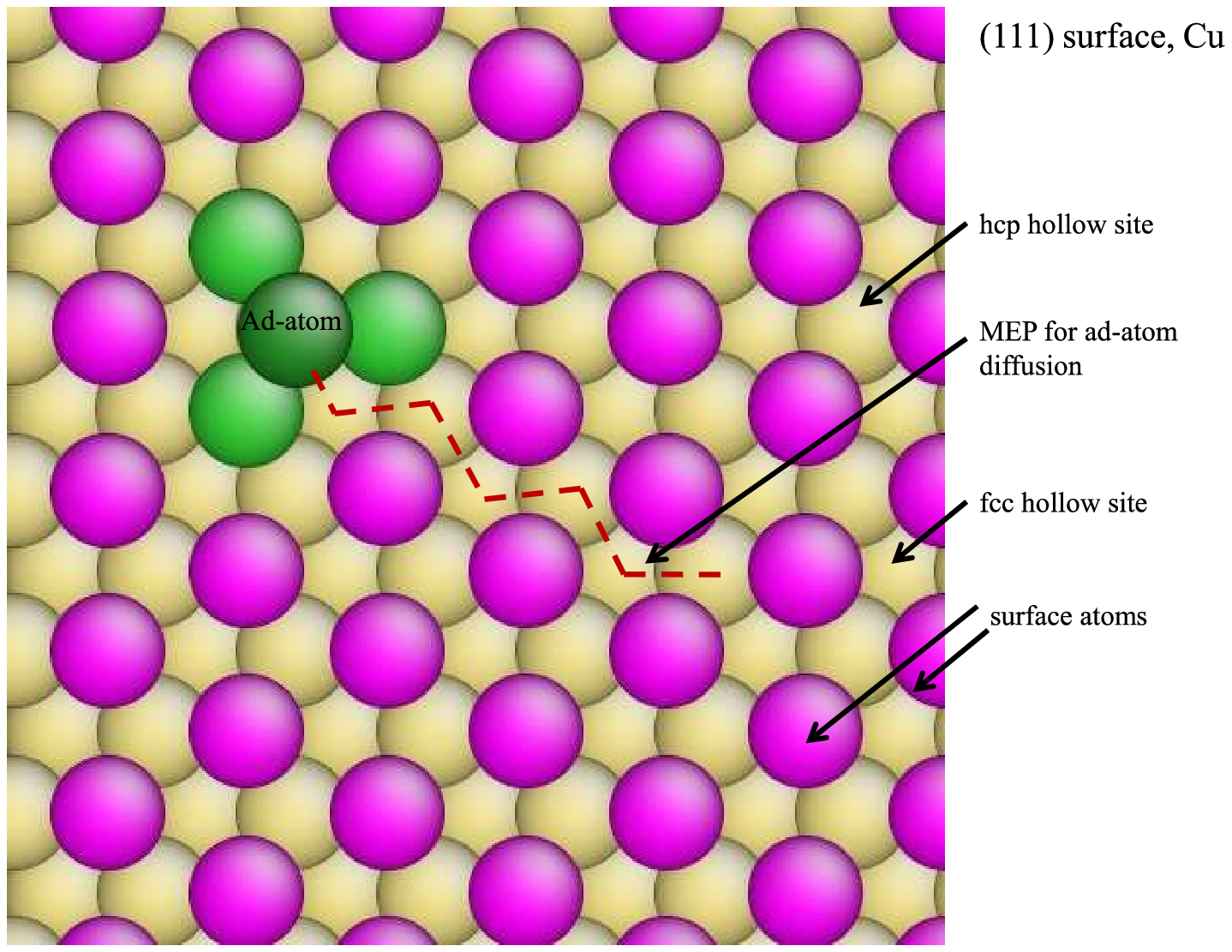}
  \caption{A snapshot of the initial structure of ad-atom on (111)
    surface of Cu sample. The MEP is shown in red dashed
    line. Coordination number (CN) coloring is used in this case. The
    atoms on the free surface have CN 9 while those near the ad-atom
    have CN 10. The ad-atom sitting on the hcp hollow site has CN 3.} 
  \label{AdatomDiffusion111surfaceCu}
\end{figure}

\end{document}